# Automated Analysis of Drawing Process for Detecting Prodromal and Clinical Dementia

Yasunori Yamada[1*], Masatomo Kobayashi[1], Kaoru Shinkawa[1], Miyuki Nemoto[2], Miho Ota[2], Kiyotaka Nemoto[2], Tetsuaki Arai[2]

[1]Digital Heath, IBM Research, Tokyo, Japan

[2]Department of Psychiatry, Division of Clinical Medicine, Faculty of Medicine, University of Tsukuba, Ibaraki, Japan

*Email: ysnr@jp.ibm.com

*Abstract*— Early diagnosis of dementia, particularly in the prodromal stage (i.e., mild cognitive impairment, or MCI), has become a research and clinical priority but remains challenging. Automated analysis of the drawing process has been studied as a promising means for screening prodromal and clinical dementia, providing multifaceted information encompassing features, such as drawing speed, pen posture, writing pressure, and pauses. We examined the feasibility of using these features not only for detecting prodromal and clinical dementia but also for predicting the severity of cognitive impairments assessed using Mini-Mental State Examination (MMSE) as well as the severity of neuropathological changes assessed by medial temporal lobe (MTL) atrophy. We collected drawing data with a digitizing tablet and pen from 145 older adults of cognitively normal (CN), MCI, and dementia. The nested cross-validation results indicate that the combination of drawing features could be used to classify CN, MCI, and dementia with an AUC of 0.909 and 75.1% accuracy (CN vs. MCI: 82.4% accuracy; CN vs. dementia: 92.2% accuracy; MCI vs. dementia: 80.3% accuracy) and predict MMSE scores with an $R^2$ of 0.491 and severity of MTL atrophy with an $R^2$ of 0.293. Our findings suggest that automated analysis of the drawing process can provide information about cognitive impairments and neuropathological changes due to dementia, which can help identify prodromal and clinical dementia as a digital biomarker.

## I. INTRODUCTION

Dementia has become a serious health and social problem. Early diagnosis of dementia, particularly in the prodromal stage (i.e., mild cognitive impairment, or MCI), is important for providing disease-modifying treatments and secondary prevention [1–4]. However, diagnosis rates remain so low that 75% of people with dementia have not been diagnosed globally [1], and the rates are particularly low for earlier stages [5, 6]. Currently, biomarkers in cerebrospinal fluid and neuroimaging as well as comprehensive neuropsychological assessments have been used for clinical diagnosis [7, 8], but these methods can be invasive, time-consuming, and expensive. In this context, easy-to-use screening tools would help identify individuals who require further examination with biomarkers and comprehensive neuropsychological assessments for diagnostic decision-making [9, 10, 11, 12].

The most common screening tests are the Mini-Mental State Examination (MMSE) [13] and Montreal Cognitive Assessment (MoCA) [14]. One of the limitations may be that both tests require administration by professionals. According to survey, primary care physicians have been reported to perceive barriers to recognizing the presence of dementia and making timely referrals to specialists [1]. In addition, 83% of clinicians pointed out the delay of screening tests due to the COVID-19 pandemic [1]. Another limitation may be their use in different languages [1]. Non-linguistic tests for cognitive screening are expected to help overcome the effect of differences in languages [1]. In summary, developing a self-administered screening tool available in various settings (e.g., primary care and at home) and to different populations would help early identification of dementia.

Drawing analysis may be a good means for developing such a screening tool, as drawing is a complex activity involving multiple cognitive function related to dementia. Pentagon-copying and clock-drawing tests are examples for the drawing tests used in screening cognitive impairments and dementia. In addition to automatic scoring in accordance with the drawing outcome, digital technologies, such as a digitizing tablets and pens, enable detailed analysis of the drawing process by recording a multitude of drawing features such as drawing speed, pen posture, writing pressure, and pauses between strokes. These drawing features have been shown to change in individuals with cognitive impairments and in patients with dementia [15–18]. For instance, slower drawing speed, increased speed variability, and longer pauses have been observed even in the prodromal stage of dementia [11, 19–21]. Combination of these drawing features could detect prodromal/clinical dementia and predict clinical scores for global cognition. For example, studies reported around 80% accuracy of a model for classifying MCI and cognitively normal (CN) individuals [11, 21], and other studies reported moderate correlations between actual and predicted MMSE or MoCA scores obtained using a regression model [22–24]. One of the latter studies also suggested the international applicability of automated drawing analysis for screening cognitive impairment

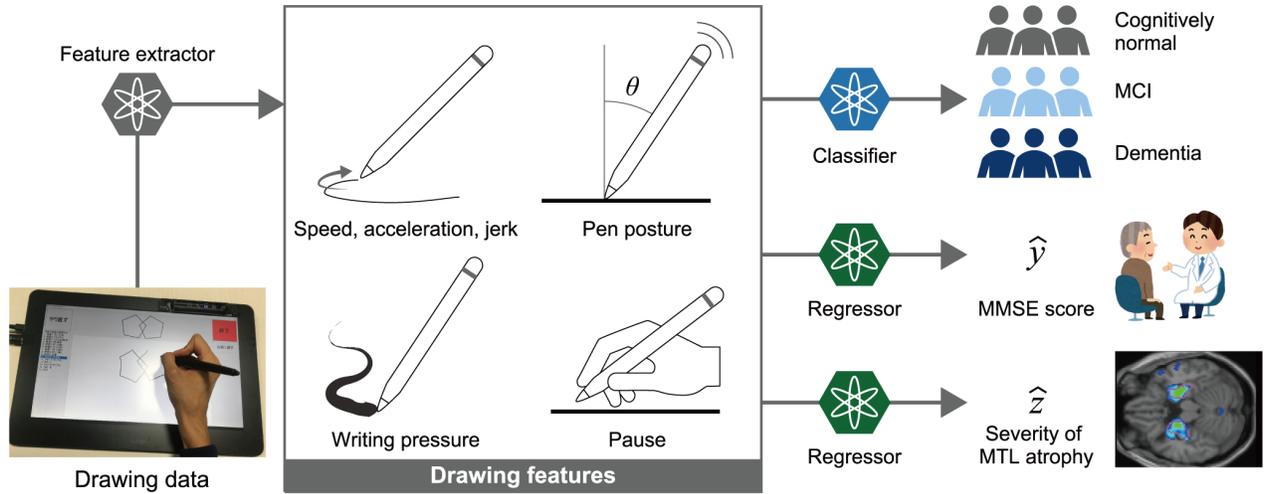

Fig. 1. Analysis overview: Capturing drawing data with a tablet; extracting 38 types of drawing features for each task; and evaluating classification and regression models for classifying the diagnostic groups of CN, MCI, and dementia, as well as for predicting MMSE scores and severity of MTL atrophy.

by investigating data collected from older adults in the United States and Japan [24].

In this study, we evaluated the feasibility of using drawing analysis for detecting prodromal and clinical dementia and for predicting cognitive and neuropathological changes related to dementia. Specifically, we collected drawing data from 145 participants (46 CN, 67 MCI, and 32 dementia) who performed five drawing tasks. We then extracted the above-mentioned drawing features. Finally, we developed models for classifying clinical diagnosis groups of CN, MCI, and dementia as well as for predicting the severity of cognitive impairments assessed by MMSE as well as the severity of neuropathological changes assessed by medial temporal lobe (MTL) atrophy.

## II. MATERIALS AND METHODS

### A. Participants

A total of 145 community-dwelling older adults in Japan were recruited (see [11] for detailed recruitment procedures). The inclusion criterion for the MCI and dementia participants was a diagnosis of MCI or dementia in accordance with standard research diagnostic criteria [7, 8, 25–27]. The CN participants did not meet any of those criteria and were age-matched to the participants in the other two groups. The participants were assessed with cognitive and clinical examinations, which resulted in twelve variables in total (Table I). The severity of MTL atrophy was expressed as a Z-score relative to healthy controls, which was evaluated with structural magnetic resonance imaging (see [11] for detailed parameters and procedures).

The study protocol was approved by the Ethics Committee, University of Tsukuba Hospital (H29-065). All examinations were conducted in Japanese with written informed consent from the participants.

### B. Drawing-data collection

During the cognitive assessments, the participants performed drawing tasks by using a Wacom tablet (Cintiq Pro 16; Fig. 1). The details of the data capturing software were presented in [24].

The following five tasks were administered for drawing-data collection: the sentence-writing and pentagon-copying items of the MMSE, Trail Making Test parts A and B (TMT-A and TMT-B) [28], and Clock Drawing Test (CDT) [29]. The sentence-writing task required writing a spontaneous sentence. The pentagon-copying task required copying a figure of intersecting pentagons. The TMT-A task required drawing lines to connect consecutive numbers. The TMT-B task required drawing lines to connect numbers and letters alternately. Finally, the CDT task required drawing an analog clock face to show 10 minutes after 10 o'clock.

### C. Drawing feature extraction

We extracted a common set of drawing features for all five tasks. Specifically, we extracted 38 types of drawing features from each task; 15 related to speed/acceleration/jerk, 8 related to pen pressure, 10 related to pen posture, and 5 related to pauses. Fig. 1 shows an overview of these features.

The features for speed, acceleration, and jerk included the median, coefficient of variation (CV) across and within strokes, and number of local extrema per unit length and time. These features were calculated from the trajectory of the pen tip. The number of local extrema was used to characterize non-smoothness. The features for pen pressure included the median, CV across and within strokes, and number of local extrema (non-smoothness) per unit length and time for pen pressure, as well as the median and CV across and within strokes for the speed of changes in pen pressure. The features for pen posture included the standard deviations across and within strokes for tilt-x and tilt-y as well as the absolute median and CV across and within strokes for the speed of changes in tilt-x and tilt-y. The

TABLE I. PARTICIPANT DEMOGRAPHICS AND COGNITIVE/CLINICAL MEASURES

| | Mean (SD) | | | p value |
|---|---|---|---|---|
| | CN (n=46) | MCI (n=67) | Dementia (n=32) | |
| Age, years | 72.3 (3.9) | 74.1 (4.5) | 74.3 (6.7) | 0.0964 |
| Sex, female, n (%) | 30 (65.2) [M] | 28 (41.8) [C] | 16 (50.0) | **0.0496** |
| Education, years | 13.2 (1.9) | 13.7 (2.6) | 12.5 (2.7) | 0.0790 |
| MMSE | 28.0 (1.6) [D] | 27.1 (1.7) [D] | 20.8 (3.9) [C,M] | **< 0.0001** |
| Frontal Assessment Battery | 13.5 (2.5) [D] | 12.7 (3.0) [D] | 9.4 (3.7) [C,M] | **< 0.0001** |
| Logical Memory IA | 11.0 (3.4) [M,D] | 7.5 (3.2) [C,D] | 3.0 (2.9) [C,M] | **< 0.0001** |
| Logical Memory IIA | 9.1 (3.0) [M,D] | 5.0 (3.5) [C,D] | 1.1 (1.9) [C,M] | **< 0.0001** |
| TMT-A | 35.7 (11.7) [D] | 43.3 (17.1) [D] | 76.7 (57.8) [C,M] | **< 0.0001** |
| TMT-B | 91.1 (40.9) [M,D] | 136.0 (78.2) [C,D] | 249.0 (70.6) [C,M] | **< 0.0001** |
| CDT | 6.7 (0.9) [D] | 6.6 (0.9) [D] | 5.8 (2.1) [C,M] | **0.0103** |
| Clinical Dementia Rating | 0.0 (0.0) [M,D] | 0.5 (0.1) [C,D] | 0.9 (0.4) [C,M] | **< 0.0001** |
| Geriatric Depression Scale | 3.2 (3.0) | 3.3 (3.3) | 3.6 (3.0) | 0.8678 |
| Barthel Index of ADL | 99.7 (1.2) [D] | 99.3 (2.3) [D] | 97.7 (5.8) [C,M] | **0.0180** |
| Lawton IADL | 7.8 (0.6) [D] | 7.3 (1.1) [D] | 5.1 (2.0) [C,M] | **< 0.0001** |
| MTL atrophy | 0.8 (0.5) [M,D] | 1.2 (0.8) [C,D] | 1.9 (1.2) [C,M] | **< 0.0001** |

Values were examined by using the chi-square test for sex and one-way analysis of variance for the other data. Significant differences between individual diagnostic groups (chi-square test, $p < 0.05$, for sex; Tukey-Kramer test, $p < 0.05$, for the other data) are marked with C, M, or D (C: different from CN; M: different from MCI; D: different from dementia).

features for pauses included the mean and CV of pause duration between drawing motions, number of drawings, pause/drawing duration ratio (the ratio of the pause duration to drawing duration), and adjusted total duration (the sum of the pause and drawing durations per unit stroke length).

### D. Machine learning analysis

We used supervised machine-learning models using drawing features to classify the clinical diagnostic groups of CN, MCI, and dementia as well as to predict MMSE scores and severity of MTL atrophy. The 190 drawing features (38 types × 5 tasks) were input into the models. We evaluated model performance with a 10 × 5 nested cross-validation procedure. To reduce overfitting, we first applied feature selection by using L1-regularized logistic regression throughout the same nested cross-validation procedure then trained classifiers and regressors on the selected features. We repeated the procedure ten times with different training and test partitions, and evaluated model performance with the area under the receiver operating characteristic curve (AUC) as well as accuracy, sensitivity, specificity, and F1 score for classifiers; and with the coefficient of determination ($R^2$), mean absolute error (MAE), and root mean square error (RMSE) for regressors.

For the classification and regression algorithm, we used logistic/linear regression with elastic net regularization, random forest, and support vector machine (SVM). The following hyperparameters were turned through cross-validation: for the logistic and linear regressors, the regularization parameter (search range: 0.1 to 1.0) and inverse of regularization strength (0.001 to 1); for the random forest, the max depth of trees (2 to 3) and max number of features (2 to 5); and for the SVM, kernel functions (linear and radial basis function), C (1 to 200), and gamma (0.0001 to 1). We set the class-weight parameter to the "balanced" mode. All other parameters were kept at the default values in scikit-learn 0.23.2.

The significance of the model performance derived from the nested cross-validation was determined through permutation tests. Specifically, we randomly permuted the participants' diagnosis labels, MMSE scores, and severity of medial temporal lobe atrophy. We then carried out the same nested cross-validation procedure previously described. We built a null distribution for the model performance by 100 permutations, and considered the performance to be significantly greater than chance when the performance value achieved with the true data was greater than 95% of that achieved during the permutations ($p < 0.05$).

## III. RESULTS

Table I summarizes the demographic and cognitive/clinical information for the participants. They comprised three diagnostic groups of 46 CN, 67 MCI, and 32 dementia participants. The dementia group comprised 25 patients with Alzheimer's disease, 6 patients with dementia with Lewy bodies, and 1 patient with frontotemporal dementia.

Through nested cross-validation, we evaluated the performance of the model using drawing features for classifying the three clinical diagnostic groups of CN, MCI, and dementia. Consequently, the three-class classification model achieved 75.1% accuracy (95% confidence interval (CI), 74.1% to 76.1%; p < 0.01; Fig. 2a) and AUC of 0.909 (95% CI, 0.899 to 0.919; p < 0.01). This AUC value was higher than an AUC of 0.861 calculated with MMSE scores as the baseline value. The classification model was based on a logistic regression with elastic regularization. The binary-classification models achieved 82.4% accuracy (81.3% sensitivity, 83.9% specificity, 84.6% F1 score, AUC of 0.908) for CN vs. MCI; 92.2% accuracy (85.3% sensitivity, 97.0% specificity, 89.9% F1 score, AUC of 0.965) for CN vs. dementia; and 80.3% accuracy (73.1% sensitivity, 83.7% specificity, 70.6% F1 score, AUC of 0.854) for MCI vs. dementia (Table II).

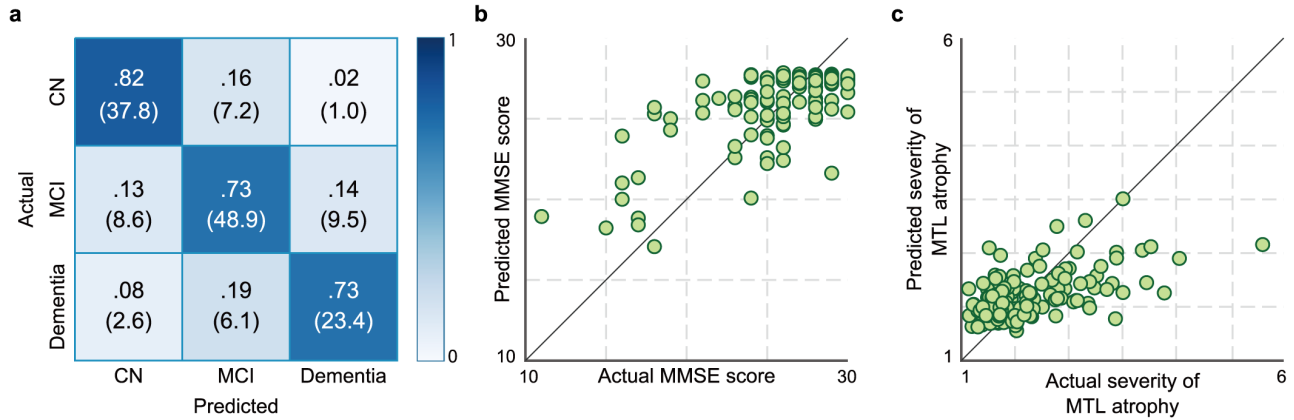

Fig. 2. Analysis results: (**a**) Confusion matrix of three-class model for classifying the three diagnostic groups. Values are presented as mean proportion (number) of participants across ten iterations; (**b**) Plot of actual and predicted scores of MMSE; and (**c**) Plot of actual and predicted severity of MTL atrophy.

We next evaluated the regression models for predicting MMSE scores and severity of MTL atrophy. The result of iterative nested cross validations indicates that the models using drawing features could predict MMSE scores with an $R^2$ of 0.491 (MAE of 1.97 and RMSE of 2.59; $p < 0.01$; Fig. 2b) and severity of medial temporal lobe atrophy with an $R^2$ of 0.293 (MAE of 0.565 and RMSE of 0.765; $p < 0.01$; Fig. 2c). The regression models were based on a random forest for MMSE and SVM with a radial basis function kernel for severity of MTL atrophy.

## IV. DISCUSSION

The results of nested cross-validation showed that the binary classifiers using drawing features could detect MCI and dementia patients with 82.4% and 92.2% accuracies, respectively. In addition, the regression model could predict the MMSE scores with an $R^2$ of 0.491 (Pearson correlation of 0.715). Previous studies on drawing analysis reported 81.2 and 93.0% accuracies for detecting MCI and Alzheimer's disease [21], Pearson correlation of 0.60 for predicting MMSE scores [23], and $R^2$ of 0.35 (Pearson correlation of 0.61) for predicting MoCA scores [24] by using features derived from a single drawing task. A few studies that exploited multi-task models reported 81.0% accuracy for detecting MCI [11] and 91.4–96.2% for detecting dementia or Alzheimer's disease [11, 30]. Compared with the performance of these state-of-the-art models, both classification and regression models in our study showed better or comparable performance. This potential improvement might be due to the combination of multifaceted drawing features derived from multiple drawing tasks, which would effectively capture diverse aspects of cognitive impairments. Our results provide the first evidence for the feasibility of using the automated analysis of the drawing process for dementia screening—a common task-neutral analytical framework was applicable both to the detection of prodromal and clinical dementia as well as the prediction of global cognition measures. We also showed that the same framework could also be useful for predicting the severity of MTL atrophy, which highlights the potential of this approach as a digital biomarker associated with neuropathological changes underlying the progression of dementia.

There is growing interest in using digitally-captured behavioral data such as drawing, speech, and motions for detecting dementia both in and outside clinical settings [9, 31]. For example, some studies used drawing and speech data collected during neuropsychological assessment for detecting MCI and dementia [22, 32], while others have used gait and daily conversational data for monitoring behavioral changes for detecting dementia [12, 33–36]. Aligning with the previous studies, our findings support the notion that automated analysis of behavioral data allows objective, continuous monitoring that would provide surrogate markers for disease prognosis or intervention effects. In addition, the task-independent nature of our analytical framework allows applying it to a variety of tasks, which may mitigate the issue of learning effects due to repetitive and frequent assessment.

TABLE II. BINARY-CLASSIFICATION PERFORMANCE FOR CLASSIFYING DIAGNOSTIC GROUPS.

| | Mean [95% Confidence Interval] | | | | |
|---|---|---|---|---|---|
| | **Accuracy (%)** | **Sensitivity (%)** | **Specificity (%)** | **F1 score (%)** | **AUC** |
| **CN vs. MCI** | 82.4 [80.8, 84.0] | 81.3 [79.7, 83.0] | 83.9 [81.8, 86.0] | 84.6 [83.1, 86.0] | 0.908 [0.900, 0.916] |
| **CN vs. Dementia** | 92.2 [91.3, 93.0] | 85.3 [83.9, 86.7] | 97.0 [95.9, 98.0] | 89.9 [88.8, 91.1] | 0.965 [0.955, 0.974] |
| **MCI vs. Dementia** | 80.3 [78.2, 82.5] | 73.1 [69.4, 76.9] | 83.7 [81.9, 85.6] | 70.6 [67.3, 73.8] | 0.854 [0.832, 0.876] |

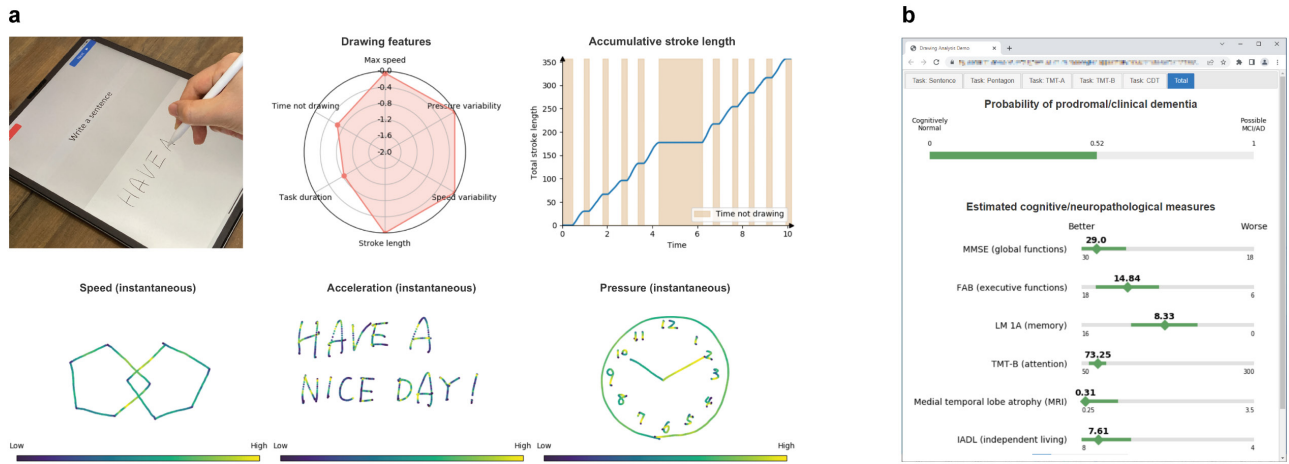

Fig. 3. Mock-up dementia screening tool based on drawing behaviors: (**a**) A user performs drawing tasks on a tablet device (iPad) while the system extracts and visualizes drawing features related to speed/acceleration, pressure, and pauses; and (**b**) A web-based report displays the outputs of machine-learning models using the extracted drawing features. The outputs include estimated probability of dementia, along with estimated cognitive and neuropathological measures to help interpretation of the screening result.

Fig. 3 shows our mock-up dementia screening tool that exploits the resultant analysis models. Analysis of drawing features would have practical advantages among digital behavioral markers for dementia screening because i) drawing is a common activity in everyday life; ii) many commercial-grade devices (e.g., mobile tablets and digital pens) are available for drawing-data collection; and iii) analysis of the drawing data may require less computing resources and raise less privacy issues compared with audio- or video-based data analysis. Previous studies have also reported the utility of drawing data for monitoring other types of diseases that are common in older adults (e.g., Parkinson's disease) [37], and many drawing tasks are language-neutral, thus promising for international use [24]. Together, our results may facilitate future efforts towards the development of worldwide applications for detecting and monitoring changes related to various types of diseases in older adults.

There were several limitations in this study. First, all analyses were carried out for all-cause dementia. Further study will be required to elucidate the association between drawing-process characteristics and neuropathological changes related to specific dementia types that would require different care strategies. Second, this was a small sample of a single center study. This might affect the generalizability of our findings, which will require further confirmation with larger samples collected in multiple centers across different populations.

In conclusion, we demonstrated the capability of automated analysis of the drawing process for detecting prodromal and clinical dementia and for predicting cognitive/neuropathological changes related to dementia. Our findings suggest the possibility of using drawing-process features as a surrogate digital biomarker for dementia, which may help develop a self-administered screening tool.

## Acknowledgment

This work was supported by the Japan Society for the Promotion of Science, KAKENHI (grant 19H01084).